\documentclass[twocolumn,showpacs,preprintnumbers,floatfix]{revtex4}

\usepackage{graphicx}
\usepackage{dcolumn}
\usepackage{bm}

\begin{document}

\title{On the Local Structure of Topological Charge Fluctuations in QCD}

\author{I.~Horv\'ath$^{a}$, S.J.~Dong$^{a}$, T.~Draper$^{a}$,
        F.X.~Lee$^{b,c}$, K.F.~Liu$^{a}$, H.B.~Thacker$^{d}$
        and J.B. Zhang$^{e}$}

\affiliation
{\centerline{$^{a}$Department of Physics, University of Kentucky, 
                   Lexington, KY 40506}\\
 \centerline{$^{b}$Center for Nuclear Studies and Department of Physics,
                   George Washington University, Washington, DC 20052} \\
 \centerline{$^{c}$Jefferson Lab, 12000 Jefferson Avenue, 
                   Newport News, VA 23606} \\
 \centerline{$^{d}$Department of Physics, University of Virginia, 
                   Charlottesville, VA 22901} \\
 \centerline{$^{e}$CSSM and Department of Physics and Mathematical Physics, 
                   University of Adelaide, Adelaide, SA 5005, Australia}}

\begin{abstract}
We consider the lattice topological charge density introduced by Hasenfratz, 
Laliena and Niedermayer and propose its eigenmode expansion as a tool to investigate 
the structure of topological charge fluctuations in QCD. The resulting effective
density is built from local chiralities studied previously. At every order of 
the expansion the density exactly sums up to the global topological charge, and 
the leading term describes the maximally smooth space-time distribution of charge 
relevant for propagating light fermions. We use this framework to demonstrate our 
previous suggestion that the bulk of topological charge in QCD does not effectively 
appear in the form of quantized unit lumps. Our conclusion implies that it is unlikely 
that the mixing of ``would-be'' zeromodes associated with such lumps is the prevalent 
microscopic mechanism for spontaneous chiral symmetry breaking in QCD. We also 
present first results quantitatively characterizing the space-time behavior 
of effective densities. For coherent fluctuations contained in spherical regions we 
find a continuous distribution of associated charges essentially ending 
at $\approx\, 0.5$. 
\end{abstract}
\maketitle


Some intriguing effects in QCD, such as the large $\eta'$ mass and the 
$\theta$-dependence, are related to vacuum fluctuations of topological 
charge~\cite{WittenUA(1)}. Understanding the {\it local} structure of topological 
charge fluctuations is thus of great interest for building a detailed picture of how these 
phenomena arise in terms of fundamental degrees of freedom. Another important phenomenon 
possibly related to topological charge fluctuations is spontaneous chiral symmetry 
breaking (S$\chi$SB). This is based on the proposition that the effective low-energy 
structure of topological charge fluctuations in QCD is such that in a typical 
configuration {\it most} of topological charge is concentrated in $N_L$ typically 
non-overlapping, four-dimensional space-time regions ${\cal L}_i$, each containing a 
sign-coherent lump of approximately unit topological charge (i.e.\ the generalization 
of the instanton liquid picture). {\it If true}, this would imply an appealing microscopic 
explanation for the origin of Dirac near-zero 
modes~\cite{ins_mixing} and hence the origin of S$\chi$SB~\cite{Banks_Casher}. 
The connection to fermions arises by associating with each lump ${\cal L}_i$ a localized 
chiral mode $\chi^i$ which ``would be'' a zeromode in the absence of other lumps, 
and a formation of a topological subspace of low-lying eigenmodes 
$\psi^i\approx \sum_j a_{ij}\chi^{j}\;,i=1,\ldots N_L$.
Verifying this ``topological mixing'' scenario has long been hindered by 
inherent difficulties in interpreting the {\it local} behavior of topological charge 
density for typical configurations of lattice-regularized theory. 
We have argued~\cite{Hor01A,Hor02A} that meaningful information can be extracted indirectly 
by studying the local 
chirality of low-lying modes. The flat behavior of the associated $X$-distribution
observed in the initial study suggested that topological mixing was not the origin of 
the near-zeromodes, and also indicated the lack of extended self-dual excitations~\cite{Hor01A}. 
However, later studies revealed the double-peaked behavior~\cite{Followup} which is 
qualitatively {\it consistent} with both these aspects, but {\it not sufficient} for their 
demonstration~\cite{Hor02A}. Being agnostic about self-duality, we observed  
in Ref.~\cite{Hor02A} that even with the
double-peak structure present, the patterns in low eigenmodes of the overlap operator 
are inconsistent with the presence of a distinctive topological subspace, and hence with 
the topological mixing scenario. In this work, 
we propose a framework generalizing the local chirality method, and use it to demonstrate 
this conclusion {\it directly}. The basis of our approach is the topological charge density 
$q_x=\frac{1}{2}\, \mbox{\rm tr} \,\gamma_5 \, D_{x,x}$~\cite{Has98A} associated with 
$\gamma_5$-Hermitian $D$ satisfying $\{D,\gamma_5\} = D \gamma_5 D$  
(Ginsparg-Wilson (GW) fermions). Note that since ${\rm tr}\,\gamma_5=0$, one can also 
write $q_x=-{\rm tr} \,\gamma_5 \, (C - \frac{1}{2}D_{x,x})$ with arbitrary constant $C$.
We choose $C=1$ (see also Ref.~\cite{Fuji99}) which guarantees that the eigenmode expansion 
of $q_x$ is properly normalized and satisfies the index theorem for arbitrary truncation 
(see below). In this case we have 
\begin{equation}
   q_x \,=\, -\mbox{\rm tr} \,\gamma_5 \, (1 - \frac{1}{2}D_{x,x})
       \,=\, -\sum_{\lambda} (1 - \frac{\lambda}{2})\, c^{\lambda}_x 
   \label{eq:15}  
\end{equation}
where $c^\lambda_x = \psi^{\lambda \,+}_x \gamma_5 \psi^{\lambda}_x$ is the local chirality 
of the mode with eigenvalue $\lambda$. For low-lying truncation $(|\lambda|\approx 0)$ 
this is effectively the sum of individual local chiralities.
Very recently, Gattringer offered the evidence for a large degree of self-duality in 
the vicinity of peaks of an eigenmode-filtered action density~\cite{Gat02A}. If the extended 
nature of self-dual regions is confirmed, then combined with our conclusions one is led to 
a picture of inhomogeneous, but essentially continuous topological charge fluctuations 
at low energy, with a significant level of self-duality in the vicinity of local maxima. 
The origin of the double-peaked behavior of local chirality would then be the {\it local} 
attraction of spinorial components by (anti)self-dual field as argued in~\cite{Hor01A}, with 
topological mixing not playing a role. In the last part of this work, we describe the first 
attempt to characterize the typical amounts of topological charge associated with coherent 
fluctuations, as well as some interesting quantitative results that emerged.

{\bf 1.} In the usual discussion of topological mixing there is a paradox rooted in the 
negativity of Euclidean topological charge density correlator at nonzero distance,
\begin{equation}
   \bigl<\,q(x) q(0) \,\bigr> \,\;\le\;\, 0 \,,\qquad |x|>0\,.
   \label{eq:05}
\end{equation}
This follows from reflection positivity and the fact that the operator $q(x)$ is  
reflection-odd~\cite{Kov,SeSt}. One consequence is that it is impossible that in typical 
configurations {\it most} of topological charge is concentrated in coherent 
four-dimensional lumps of finite physical size (e.g. instantons). Indeed, if this was 
the case and the typical size of such lumps was $r_c$, then the average correlator would 
be positive over $r_c$, thus contradicting (\ref{eq:05}). This means that lumps ${\cal L}_i$
cannot be identified in $q(x)$. 
 
These considerations do not rule out topological mixing as an effective low-energy scenario 
for S$\chi$SB, but emphasize the necessity of some short-distance filtering when studying 
these issues. This can be achieved by replacing the local operator $q(x)$ by a {\it nonlocal} one. 
While the use of nonlocal operators is generally unacceptable, here they will only serve as a filter
to smear the singular contact term in (\ref{eq:05}) into a finite positive core. The dominance 
of coherent structures is then possible. Following up on our reasoning in Refs.~\cite{Hor01A,Hor02A}, 
we propose that the low energy truncation of the Dirac eigenmode expansion for $q(x)$ (see below) 
can serve as the physically motivated nonlocal operator needed. We stress the physical motivation 
because, for the problem at hand, the underlying issue is whether the light fermion effectively 
feels a collection of coherent unit lumps as it moves through the vacuum, or there is some other
dynamics governing its propagation. This is to be decided by the fermion.

The desired fermion filtering can be realized starting from lattice-regularized theory. 
Among the lattice topological charge density operators considered, the ones associated
with GW fermionic kernels are perhaps theoretically most appealing~\cite{Has98A}.
These are constrained so that they sum up to integer global charge defined by counting the 
exact zeromodes of the GW kernel used. The continuum gauge-fermion correspondence 
(index theorem) for smooth gauge fields is thus extended to the lattice by 
construction~\cite{Has98A}. Before defining the filtered densities starting from 
(\ref{eq:15}), it is useful to recall that the spectrum of $D$ contains $N_0$ zeromodes with 
global chirality $+1$ ($N_0^+$) or $-1$ ($N_0^-$), $N_2$ modes at $\lambda=2$ with global 
chirality $+1$ ($N_2^+$) or $-1$ ($N_2^-$), and $N_p$ pairs of complex modes with zero global 
chirality. The dimension of Dirac space is $N=N_0+N_2+2*N_p$, and the topological charge is 
$Q=\sum_x q_x = N_0^- - N_0^+ = N_2^+ - N_2^-$. 
We now associate with $q_x$ the set of related densities 
\begin{equation}
   q_x^{(k)} = -\sum_{i=1}^{N_0} c^{0,i}_x\,
               -\sum_{j=1}^k (2 - \mbox{\rm Re}\, \lambda_j)\, c^{\lambda_j}_x 
   \label{eq:20}      
\end{equation} 
representing the truncated eigenmode sum that includes all zeromodes and the $k$ lowest-lying 
complex pairs. Complex eigenvalues $\lambda_j$ enter the sum ordered on the upper 
(lower) half of the spectral circle. Eigenmode-filtered densities $q^{(k)}$ are real
and $\sum_x q_x^{(k)}=Q=N_0^- - N_0^+$ thus leading to identical global fluctuations 
satisfying the index theorem. We have
$q_x^{(N_p)}=q_x$ and since $q^{(0)}$ can be identically zero for $Q=0$ configurations,
we consider $q^{(1)}$ to be the leading order in the expansion. The infrared eigenmodes of 
$D$ are significantly smoother than the underlying gauge field~\cite{Hor02A}, and $q^{(1)}$ 
is expected to be maximally smooth. As $k$ increases, $q^{(k)}$ gradually incorporates more 
short-distance structure. This does not mean that there is a strict new cutoff present 
in filtered densities. However, ultraviolet fluctuations irrelevant for propagation of 
light quarks are filtered out.

\begin{table}[t]
  \centering
  \begin{tabular}{cccccc}
  \hline
  \multicolumn{1}{c}{ensemble}  &
  \multicolumn{1}{c}{$\beta$}   &
  \multicolumn{1}{c}{$a$ [fm]}  &
  \multicolumn{1}{c}{$V$}       &
  \multicolumn{1}{c}{configs}   &
  \multicolumn{1}{c}{eigenpairs}\\
  \hline
  ${\cal E}_1$ & 6.00  & 0.093 & $14^4$  & 12  & 2\\
  ${\cal E}_2$ & 6.20  & 0.068 & $20^4$  & 8   & 2\\
  ${\cal E}_3$ & 6.55  & 0.042 & $32^4$  & 5   & 2\\
  ${\cal E}_4$ & 5.91  & 0.110 & $12^4$  & 6   & 9\\
  ${\cal E}_5$ & 6.20  & 0.068 & $20^4$  & 3   & 10\\
\hline
\end{tabular}
\caption{Ensembles of Wilson gauge configurations.}
\vspace*{-0.18in}
\label{table:1} 
\end{table}

\begin{figure}
\vspace*{0.08in}
\vbox{
\includegraphics[width=5.5cm]{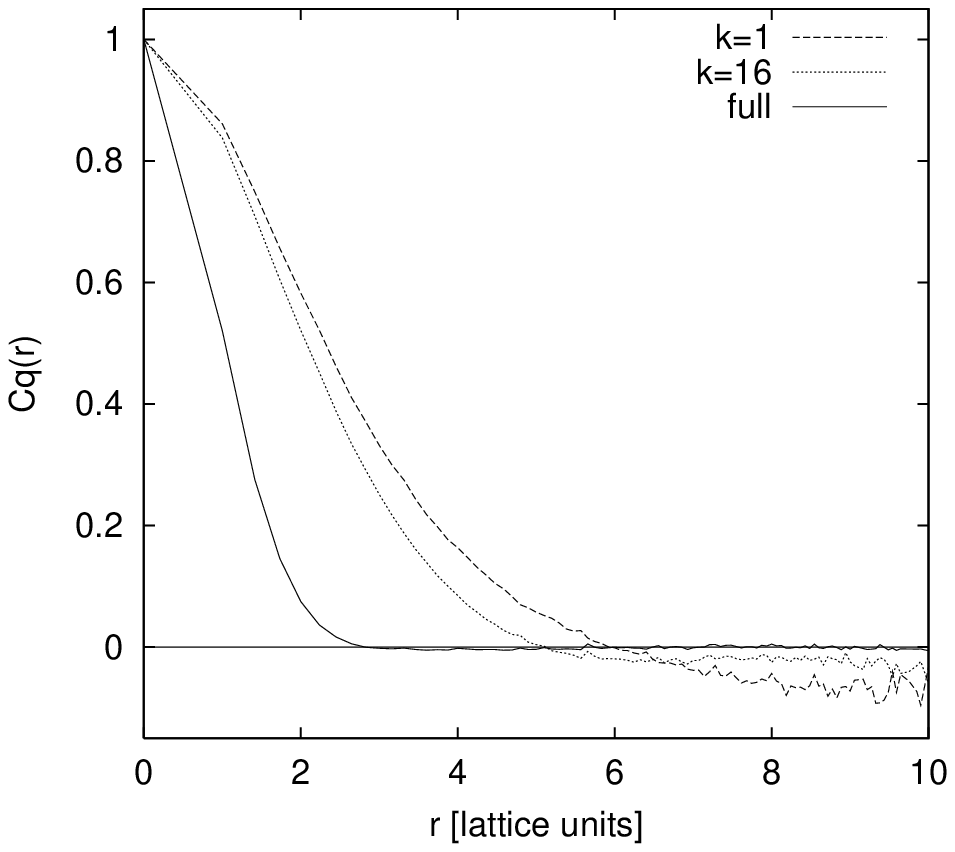}
\vskip 0.06in
\centerline{
\includegraphics[width=4.7cm]{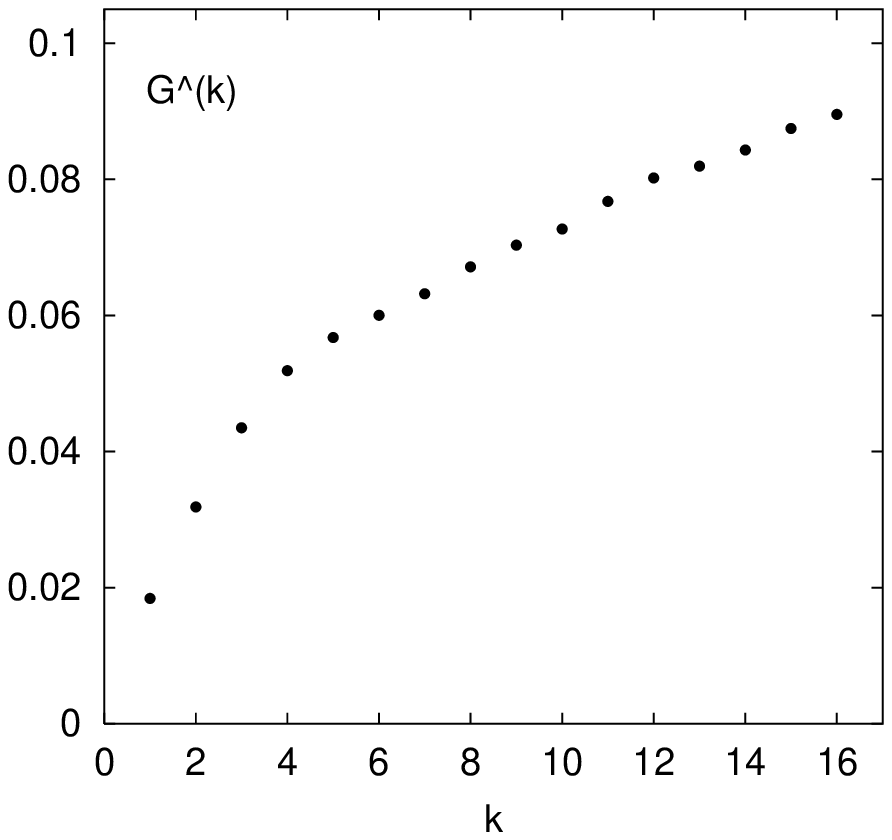}
\hspace*{-0.12in}
\includegraphics[width=4.7cm]{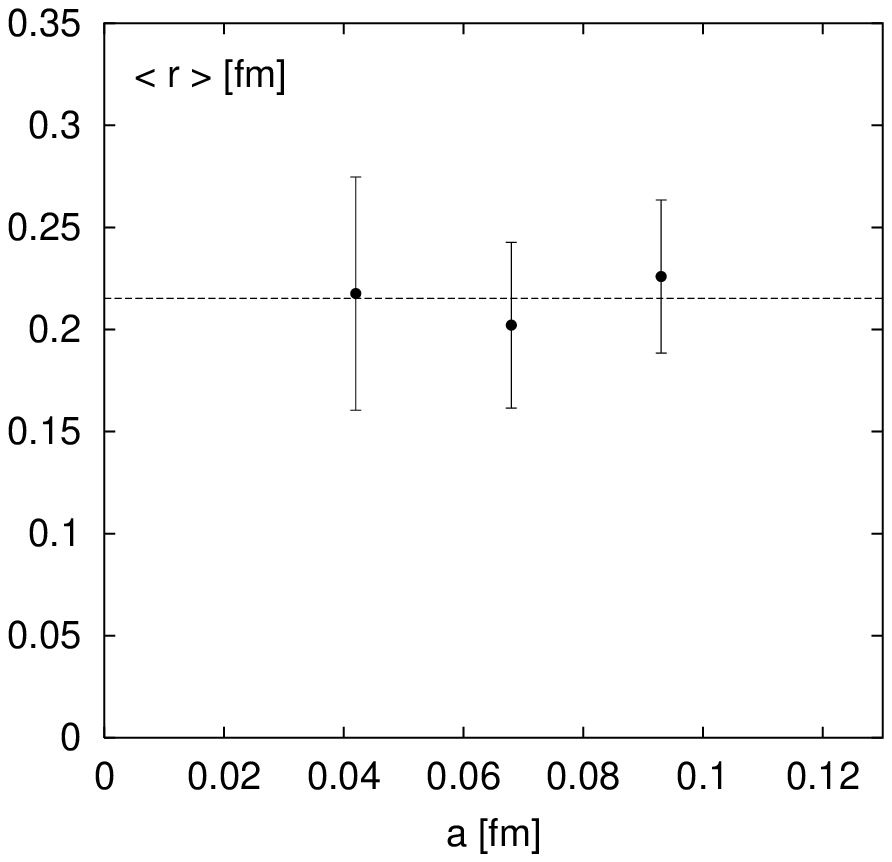}}
}
\vspace*{-0.05in}
\caption{Top: $C_q$, $C_{q^{(1)}}$ and $C_{q^{(16)}}$ for configuration ${\cal C}_2$ 
($Q=0$) from ${\cal E}_4$. Left: Roughness $G^{(k)}$ of $q^{(k)}$ for the same configuration. 
For the full density, $G^{(N_p)}=0.87$. Right: Size of the positive core of the average 
$C_{q^{(2)}}$ correlator for ensembles ${\cal E}_1$-${\cal E}_3$. Sommer parameter was
used to set the scale.}
\label{fig:smooth}
\vspace*{-0.20in}
\end{figure}

We now use the overlap Dirac operator~\cite{Neu98BA} (see details in Ref.~\cite{Hor02A})
on Wilson gauge backgrounds (see Table~\ref{table:1}) to demonstrate the basic properties 
of filtered densities discussed above. We have calculated the full density $q$ for 
configuration ${\cal C}_2$ from ensemble ${\cal E}_4$. In Fig.~\ref{fig:smooth} we show 
its correlator $C_q(r)$ (normalized at the origin) and compare it to $C_{q^{(1)}}$ and 
$C_{q^{(16)}}$ for filtered densities. Note that $C_q$ has a very short-ranged positive 
core as expected from~(\ref{eq:05}). This provides indirect evidence that the locality of 
the non-ultralocal operator $q$ is quite good. As for the correlators $C_{q^{(k)}}$, their 
range is evidently larger. The shape stabilizes at about $k=6$ and changes very slowly from 
then on. To further characterize the roughness of filtered densities we calculate
\begin{equation}
     (\,G^{(k)}\,)^2 \,\equiv\, \sum_{x,\mu}\, (\,q^{(k)}_{x+\mu}-q^{(k)}_{x}\,)^2
     \label{eq:25}
\end{equation}
and plot $G^{(k)}$ as a function of $k$. As expected, $q^{(k)}$ becomes rougher
as $k$ increases. While the {\it physical size} of the positive core in $C_q$ is 
expected to go to zero in the continuum limit, this is not necessarily so for $C_{q^{(k)}}$. 
To see that, we have calculated $C_{q^{(2)}}$ for ensembles ${\cal E}_1$-${\cal E}_3$. 
The size of the positive core in the average correlator was determined as $<r>$ over 
the probability distribution given by $C_{q^{(2)}}(r)$ in the range from zero up to the 
maximal distance where $C_{q^{(2)}}(r)$ is manifestly positive (with errors taken into 
account). As can be seen in Fig.~\ref{fig:smooth}, the size scales well, indicating that 
if the dominance of coherent four-dimensional structures in $q^{(2)}$ can be established, 
these structures can survive the continuum limit. 

{\bf 2.} Introduction of fermion-filtered densities allows us to study the relevance 
of topological mixing {\it consistently}. If the topological subspace of low-lying 
modes $\psi^i\approx \sum_j a_{ij}\chi^{j}\,,\,i=1,\ldots,N_L$ is formed, then 
\begin{equation}
    \sum_{x\in {\cal L}_i} \sum_{j=1}^{N_L}\, \psi_x^{j+}\gamma_5 \psi_x^j
     \,\approx\, s_i\,, \quad
     \gamma_5 \chi^i = s_i \chi^i\,,\quad
     s_i=\pm 1\,.
     \label{eq:30}
\end{equation}
This implies that the structure of quantized unit lumps ${\cal L}_i$ will be revealed 
in $q^{(k)}$ when all modes belonging to the topological subspace are included. Indeed, 
from equation (\ref{eq:30}) we have 
\begin{equation}
    \sum_{x\in {\cal L}_i} \, q^{(k)}_x \,\equiv\, Q_i \approx \pm 1\,,\quad
    2k \approx N_L-N_0
    \label{eq:35}
\end{equation}
since $\mbox{\rm Re}\,\lambda \approx 0$ for modes in topological subspace.
Moreover, the value of pure gauge topological susceptibility 
($\approx\! 1\, \mbox{\rm fm}^{-4}$) constrains $N_L$ in lump-dominated configurations of 
volume V fm$^4$ to be $N_L\approx V$. 
This implies that the lumpy structure should typically be saturated in $q^{(k)}$ with 
$k\:\raisebox{-0.8ex}{$\stackrel{\raisebox{-1.5ex}{$\textstyle
<$}}{\sim}$}\:V/2$. Consequently, $q^{(2)}$ is expected to be well sufficient for all 
ensembles in Table~1. 


\begin{figure}
\vspace*{-0.10in}
\includegraphics[height=5.1cm]{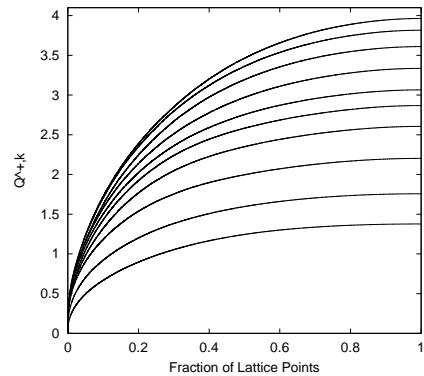}
\vspace*{-0.08in}
\caption{The function $Q^{+,k}(f)$ for configuration ${\cal C}_2$ ($Q=1$) of 
ensemble ${\cal E}_5$. The lowest curve corresponds to $k=1$.}
\vspace*{-0.00in}
\label{fig:accumq}
\end{figure}

\begin{figure}
\vspace*{-0.12in}
\includegraphics[height=5.1cm]{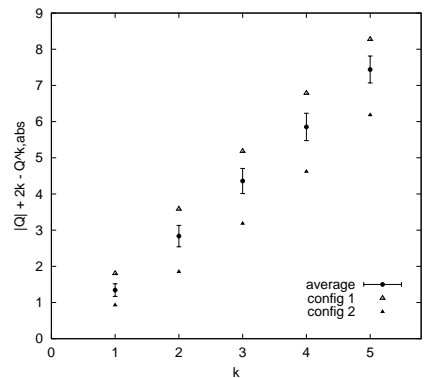}
\vspace*{-0.12in}
\caption{$|Q| + 2k - Q^{k,abs}$ for ensemble ${\cal E}_4$.}
\vspace*{-0.12in}
\label{fig:delum}
\end{figure}

We now ask whether the subset $\cup_i {\cal L}_i$ of the lattice containing most 
of topological charge can be identified using $q^{(k)}$. A simple way to proceed is to 
order lattice points by the magnitude of $q_x^{(k)}$ and compute the running sum of 
positive ($Q^{+,k}(f)$) and negative ($Q^{-,k}(f)$) charge as the fraction $f$ of the 
highest points included increases. Functions $Q^{+,k}$ ($Q^{-,k}$) should stabilize to 
a constant ($\approx$ integer-valued) plateau at a well-defined value of $f$ corresponding 
to fraction of volume occupied by $\cup_i {\cal L}_i$. In Fig.~\ref{fig:accumq} we show 
the behavior of $Q^{+,k}$ for configuration ${\cal C}_2$ ($Q=1$) from ${\cal E}_5$. We find 
no sign of plateaus for any $k$, and nothing special happens around $k=2$. Such behavior 
is characteristic for {\it all} configurations from the ensembles in Table~\ref{table:1}. 
The same conclusion applies to functions $Q^k(f)$ monitoring total charge. The smooth 
monotonic behavior exhibited in Fig.~\ref{fig:accumq} excludes the possibility that 
the bulk of topological charge is effectively concentrated in a small subvolume 
$\cup_i {\cal L}_i$ of typically isolated lumps. 

For another quantitative test, consider a configuration with topological charge $Q$ and 
{\it assume} it is dominated by $|Q|+\xi$ unit lumps (antilumps) and $\xi$ antilumps 
(lumps). If the number of zeromodes is minimal, i.e.\ $|Q|=N_0$ (true for all our configurations), 
then the dimension of the topological subspace is $N_L=|Q|+2\xi$ and we should have 
$\sum_x |q^{(\xi)}_x|\equiv Q^{\xi,abs}\approx N_L = |Q|+2\xi$. In fact, the 
quantity $|Q| + 2k - Q^{k,abs}$ would be close to zero for all $k\le\xi$ (since each mode
in the topological subspace contributes approximately unity to $Q^{k,abs}$), and it 
would start to increase rapidly for $k>\xi$. In a theory where the topological mixing scenario 
is relevant, monitoring the $k$-dependence of $|Q| + 2k - Q^{k,abs}$ could thus serve as a procedure 
for determining the dimension of the topological subspace for a given configuration. We have computed 
this $k$-dependence for all our configurations. The minimal value is always achieved at $k=1$ and 
is of order one rather than close to zero. Moreover, in ensembles ${\cal E}_4$, ${\cal E}_5$, where 
a wide range of $k$ is available, we observe monotonic (linear) increase for every configuration.
This robust behavior is illustrated in Fig.~\ref{fig:delum}, confirming again that there 
are no signs of a distinctive topological subspace and hence no signs of dominance 
by unit lumps.

\begin{figure}
\vspace*{-0.12in}
\includegraphics[height=5.1cm]{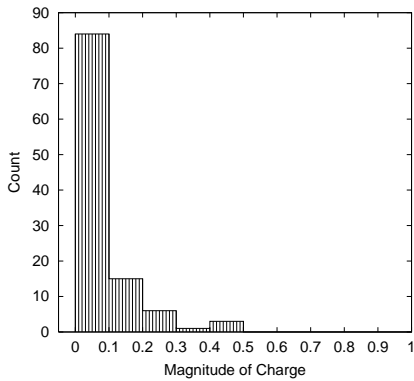}
\vspace*{-0.12in}
\caption{Distribution of charges in coherent fluctuations for ensemble ${\cal E}_4$.}
\vspace*{-0.15in}
\label{fig:qdist}
\end{figure}

{\bf 3.} The above arguments do not depend on specific properties of ${\cal L}_i$, 
such as their shape, volume or particular field content. The results show that the topological 
charge is effectively carried by the bulk of the lattice, which is inconsistent with dominance 
by unit quantized lumps. At the same time, we find inhomogeneous behavior with noticeable peaks 
in $q^{(k)}$, accompanied by visible coherence (see $C_{q^{(k)}}$ in Fig.~\ref{fig:smooth}). 
As a first step toward 
understanding this structure we now provide a simple characteristic of the typical values 
of charge associated with such coherent fluctuations. Given an arbitrary density $q$ we consider 
the sets ${\cal I}^l$ of {\it centers} of coherent behavior with lattice resolution $\sqrt l$, 
namely
\begin{displaymath}
   {\cal I}^l \,\equiv\, \{\,x \;:\; |q_y| < |q_x|,\, q_x q_y > 0 \;\;
   \forall y\;;\;0<|x-y|^2\le l\,\}\,.  
\end{displaymath} 
The elements of ${\cal I}^l$ are local maxima of $|q|$ over distance $\sqrt l$, for which the 
sign of $q$ is coherent over at least the same distance. Obviously, 
${\cal I}^1 \supset {\cal I}^2 \supset {\cal I}^3 \supset \ldots$. To every $x\in {\cal I}^l$ 
we assign a radius $R_x$ defined as the maximal distance from $x$ over which the density 
is still coherent, i.e. $R_x \ge \sqrt{l}$. The charge $Q_x$ is also assigned by summing the 
density over the sphere of radius $R_x$ centered at $x$. However, to be able to interpret $Q_x$ 
as a charge corresponding to an individual fluctuation, there should typically be no other 
centers within $R_x$. This allows for fixing the resolution in a self-consistent way by choosing 
the smallest $l$ with this condition satisfied. 

To carry out the above procedure meaningfully for $q^{(k)}$, one should work in the range 
of $k$ where coherent fluctuations and their charges are relatively stable with respect to a 
change of $k$. For our ensembles, this is best satisfied for ${\cal E}_4$ where $q^{(9)}$ is 
available. In physical terms, this corresponds to including eigenmodes with imaginary parts up 
to $\approx$ 500 MeV. The fraction of overlapping centers drops dramatically (to 5 \%) at $l=3$, 
which is the value we have used. The corresponding distribution of charges for non-overlapping 
centers is shown in Fig.~\ref{fig:qdist}. A property which is insensitive to the choice of $k$ 
and $l$ is that the distribution effectively ends at about $0.5$. This appears to hold also in 
the case of full density $q_x$ (with only two configurations available). Interestingly, this 
behavior might be compatible with the presence of center vortices in the QCD vacuum as the recent 
discussion of topology in the field of an idealized vortex suggests~\cite{Rein01A} (The possible 
manifestation of center vortices in topological charge fluctuations was also recently discussed 
in Ref.~\cite{Cor02}.).
 
To summarize, we have proposed that the low-energy behavior of topological charge density 
can be studied by using a suitable non-local realization of $q(x)$ as a filter for short-distance 
fluctuations. If one is interested in the aspects of topological charge affecting the low-momentum 
propagation of light quarks, then the appropriate non-local realization is naturally available 
through the low-eigenmode expansion of $q(x)$. Such fermion filtering can be explicitly realized 
starting from lattice-regularized theory. We have proposed the expansion of $q_x$ associated with 
GW fermions as an ideal tool for this purpose. This provided us with a consistent framework 
to test whether the propagation of light fermion is {\it effectively} driven by 
the dominance of unit topological lumps in the QCD vacuum. We find that this is not the case, 
implying that the picture of S$\chi$SB based on the mixing of corresponding topological 
``would-be'' zeromodes is not accurate. We emphasize that we have not ruled out the logical 
possibility that some unit quantized structures with well-defined boundaries can occur. 
However, we have shown that the {\it bulk} of topological charge does not come in that form. 
A first attempt to characterize the inhomogeneous nature of topological charge fluctuations 
in filtered densities resulted in very interesting results indicating that patterns of local 
behavior can provide us with detailed information about dynamically important structures 
in the QCD vacuum.

\vspace*{-0.30in}

\begin{acknowledgments}
\vspace*{-0.12in}
The support by U.S. Department of Energy grants
DE-FG05-84ER40154, DE-FG02-95ER40907, DE-FG02-97ER41027 and DE-AC05-84ER40150 is acknowledged.
\end{acknowledgments}

\end{document}